\documentclass[aps,preprint]{revtex4}%
\usepackage{amsfonts}
\usepackage{amsmath}
\usepackage{amssymb}
\usepackage{graphicx}%
\begin{document}
\title[Power laws of complex systems]{Power laws of complex systems from Extreme physical information}
\author{B. Roy Frieden}
\affiliation{College of Optical Sciences, Univ. of Arizona, Tucson, Arizona 85721}
\author{Robert A. Gatenby}
\affiliation{Radiology Dept., Arizona Health Sciences Ctr., Tucson, Arizona 85726}

\begin{abstract}
Many complex systems obey allometric, or power, laws $y=Yx^{a}$. \ Here
$y\geq0$ is the measured value of some system attribute $a$, $Y$ $\geq0$ is a
constant, and $x$ is a stochastic variable. \ Remarkably, for many living
systems the exponent $a$ is limited to values $n/4,$ $n=0,\pm1,\pm2...$ Here
$x$ is the mass of a randomly selected creature in the population. These
quarter-power laws hold for many attributes, such as pulse rate ($n=-1$).
\ Allometry has, in the past, been theoretically justified on a case-by-case
basis. \ An ultimate goal is to find a common cause for allometry of all types
and for both living and nonliving systems. \ The principle $I-J=extrem.$of
Extreme physical information (EPI) is found to provide such a cause. \ It
describes the flow of Fisher information $J\rightarrow I$\ from an attribute
value $a$ on the cell level to its exterior observation $y.$ Data $y$ are
formed\ via a system channel function $y\equiv f(x,a)$, with $f(x,a)$ to be
found. \ Extremizing the difference $I-J$ through variation of $f(x,a)$
results in a general allometric law $f(x,a)\equiv\ y=Yx^{a}$. \ Darwinian
evolution is presumed to cause a second extremization of $I-J,$ now with
respect to the choice of $a$. \ The solution is $a=n/4,$ $n=0,\pm1,\pm2...,$
defining the particular powers of biological allometry. \ Under special
circumstances, the model predicts that such biological systems are controlled
by but two distinct intracellular information sources. \ These sources are
conjectured to be cellular DNA and cellular transmembrane ion gradients

\end{abstract}
\volumeyear{2005}
\volumenumber{ }
\issuenumber{ }
\eid{ }
\date{}
\received[Received text]{}

\revised[Revised text]{}

\accepted[Accepted text]{}

\published[Published text]{}

\startpage{1}
\endpage{25}
\maketitle

\section{FISHER INFORMATION}

\ Fisher information $I$\ is defined by the following problem of estimation.
\ An unknown attribute value $a$ of a system is measured as a data value $y.$
\ The system's likelihood law $p(y|a)$ is known. \ How well can $a$ be
estimated on the basis of $y$? \ Assume that the estimate is to be unbiased.
\ Then $the$ $minimum$ $possible$ $mean-squared$ $error$ $of$ $any$ $such$
$estimate$ $is$ $1/I$ [1],[2], where%

\begin{equation}
I\equiv I(a)\equiv\left\langle \left[  \frac{\partial}{\partial a}\ln
p(y|a)\right]  ^{2}\right\rangle _{y},~~I\geq0. \tag{1a}%
\end{equation}
The notation $<~>_{y}$ means the expectation over all possible data values $y$
(see also Eq. (6)). \ As indicated, $I$ is positive by construction, and the
dependence upon the datum $y$ is averaged out, leaving only a dependence
$I(a)$\ upon the attribute value $a.$ Thus the information depends upon all
possible $y$, and is a system property. \ Also note the reasonable tendencies:
As the information becomes larger the minimum error $1/I$\ becomes smaller;
etc. \ In its multidimensional form [3],[4], $I$ measures system complexity as
well (Sec.\ IV).

Information $I(a)$ is that in $the$ $data$. \ This is to be distinguished from
a second type of Fisher information, denoted as $J(a)$,\ which is the amount
that originates at $the$ $source$ of the data. \ Thus, any observation results
from a flow \
\begin{equation}
J(a)\rightarrow I(a) \tag{1b}%
\end{equation}
of information from source to data.\ This flow of information is the basis for
the EPI variational approach (2) discussed further below.

Knowledge of the likelihood law $p(y|a)$ allows $I$ [by Eq. (1a)] and hence
the minimum possible error, to be known. \ This minimum error can be compared,
as a benchmark, with that expected from any proposed estimation approach.
$\ This$ $has$ $been$ $the$ $traditional$ $use$ $of$ $information$ $I$ $since$
$about$ $1922$ [1],[2].

However, the information $I(a)$ is currently being used in a different way --
\ to determine the scientific law that is obeyed by a complex system. \ The
law defines the system through the probability density functions (PDFs)
$p(y|a)$ or probability amplitudes that characterize the system. The system is
of a general nature (physical, biological, economic, etc.). \ For this purpose
Fisher informations\ $I(a)$, $J(a)$ are used in the principle of "Extreme
physical information" or EPI. \ This is a variational principle (Sec. II)
whose output is the sought law that governs the system [3],[4]. An example is
the Schrodinger wave equation governing the probability amplitudes of a
quantum-level system.

\section{SYSTEM AS INFORMATION CHANNEL}

Consider a system consisting of a source effect specified by an attribute
value $a$, an instrument \ for observing it (via probe particles), and the
output space consisting of a datum $y$ from the instrument. \ This defines an
"information channel". \ \ Such a system is defined by its likelihood law
$p(y|a)$ and any relations among its variables $y,a$. \ The general aim of EPI
is to determine the likelihood law and these relations. \ To facilitate
finding these, the observing instrument is assumed to be ideal and noise-free.

In general, information is lost in transition from source level $J$ to data
level $I.$\ However, data tend to at least approximate their ideal (system)
values, so that the loss of information tends to be minimal. \ Indeed,
otherwise the act of observation would be pointless. \ Hence the
principle\ \ \ \ \ \ \ \ \ \ \ \ \ \ \ \ \ \ \ \ \ \
\begin{align}
I-J  &  =extremum,\ \text{where}\ I\equiv I(a),~J\equiv J(a),~~~J\geq
0,~I=\kappa J,~\tag{2}\\
~\kappa &  =\kappa(a)=const.,~0\leq\kappa\leq1.\nonumber
\end{align}
This is called the principle of Extreme physical information (EPI). \ By (2),
$\kappa(a)\equiv I(a)/J(a)$ is a function of $a$ and is assumed to be
constant. \ For further details on origins of the EPI\ principle, see the
article [3] or the books [4]. \ \ \ \ \ \ 

\section{ALLOMETRIC SCALING LAWS}

$Note$: We use the terminology "allometry", "allometric scaling laws",
"scaling laws" and "power laws" interchangeably for Eqs. (3a,b).

\textbf{A. General allometric laws}

Allometric power laws have a general form%

\begin{equation}
\text{\textbf{y}}_{n}=\text{\textbf{Y}}_{n}x^{a_{n}},~a_{n}=\text{const.}%
,~~\ n=0,\pm1,\pm2,...,\pm N,~\ \ \ \text{\textbf{y}}_{n},\text{\textbf{Y}%
}_{n}\geq0,~~~\ 0<x<\infty. \tag{3a}%
\end{equation}
\ These are simple power laws, where each member of an attribute class $n$
obeys the same power law. \ The laws describe, to a good approximation,
certain living and nonliving systems. \ In general $n$ defines an $n$th class
of observed attributes \textbf{y}$_{n}\equiv y_{n1},y_{n2},...,y_{nK_{n}}$ of
a system. \ Also, \textbf{Y}$_{n}\equiv Y_{n1},Y_{n2},...,Y_{nK_{n}}$ is a
corresponding vector of $constants$, and$\ K_{n}$ is the number of attributes
in the class. \ Thus there is a total of $K\equiv\sum_{n}K_{n}$ attributes
over all classes. Quantity $x$ is an $independent$ $variable$ of a system that
is sampled for one of these attributes. \ The powers $a_{n}$ in (3a)\ are
empirically-defined values of the various attributes and are regarded as ideal
identifiers of these. \ The $a_{n}$ are generally dimensionless numbers such
as 2/3, 0.7, etc. \ \ Current approaches for explaining general allometry are
"self organized criticality" (SOC) [5], Lande's model [6], the scale-free (SF)
network property [7], and others [5]. \ 

\textbf{B. Biological allometric laws}

Likewise there are\ many $living$ systems that obey allometry [8]-[17],%

\begin{align}
y_{nk}  &  =Y_{nk}x^{a_{n}},~~~a_{n}=n/4,~~\ n=0,\pm1,\pm2,...,\pm
N,~~\tag{3b}\\
k  &  =1,2,...K_{n},~~~~y_{nk},Y_{nk}\geq0.\nonumber
\end{align}
Here $x$ is specifically the mass of the organism and the dimensionless powers
\ $a_{n}$\ identify attributes of the organism. \ Remarkably, $each$ $power$
$is$ $always$ $some$ $integer$ $multiple$ $of$ $1/4$. \ \ Why this should
generally be so, both within individuals and across different species, is a
great mystery of biology [9], and is addressed by this paper. \ Living systems
have "extraordinary" complexity, and in fact are reputed to be "the most
complex and diverse physical system[s] in the universe" [9]. \ This suggests
that EPI -- which applies to complex systems --\ is applicable to derivation
of these allometric laws.

Note that the same power $n/4$\ describes all $K_{n}$\ members of an $n$th
class of attributes. \ For example, the class $n=-1$ has currently $K_{-1}=2$
known attributes, consisting of the observed heart rate and observed RNA
concentration of the organism. \ The dynamic range of mass values $x$ in (3b)
by definition includes mass values that extend from some (unknown) $very$
$small$ and finite value to some (unknown) $very$ $large$ and finite value.
\ Indeed, for the attribute $n=3$ of metabolic rate, the dynamic range of $x$
over which (3b) is known to hold currently exceeds $27$ orders of magnitude [8],[9].

Allometric laws (3b) describe both $individual$ and $collective$ properties of
animals. \ Some examples are as follows. The attribute class $n=-1$ mentioned
above obeys a power $a_{-1}=-1/4$. \ The class $n=3$ has $K_{3}=1$ member
defining metabolic rate and obeys a power $a_{3}=3/4.$ Eq. (3b) even holds for
a class $n=0$, i.e. where the attributes do not vary with mass. \ An example
is hemoglobin concentration in the blood, which does not vary appreciably with
body (or mass) size. \ Other examples [9] are "metabolic rate, life span,
growth rate, heart rate, DNA nucleotide substitution rate, lengths of aortas
and genomes, tree height, mass of cerebral gray matter, density of
mitochondria, and concentration of RNA." \ \ This list of $K=11$ attributes
only scratches the surface.

\textbf{C. On models for biological allometry}

Although many biological attributes obey the quarter-power law (3a,b), many
$do$ $not$ (e.g., attributes that are the square roots of attributes that do).
\ Nevertheless, many models exist for explaining cases of biological allometry
[6],[8],[9]-[13], as conveniently summarized in [9].

However [9], these models are lacking in not providing a $unified$ $approach$
to calculating the attributes. \ Instead, they were "designed almost
exclusively to understand only the scaling of mammalian metabolic rates, and
and do not address the extraordinarily diverse, interconnected, integrated
body of scaling phenomena across different species and within individuals ...
Is all life organized by a few $fundamental$ $principles$?"

A general approach would also have to predict circumstances where allometry
will $not$ occur. \ \ A step in this direction is to find a model that
establishes necessity for $allometry$ $of$ $all$ $types$, biological and
nonliving. \ That is, it would show that:
\begin{equation}
If\text{ }a\text{ }given\text{ }attribute\text{ }obeys\text{ }the\text{
}model,\text{ }then\text{ }it\text{ }must\text{ }obey\text{ }allometry.
\tag{3c}%
\end{equation}
We next form such a model. \ This dovetails with the use of EPI, which
likewise requires a model.

\section{PRIOR KNOWLEDGE}

The high degree of $complexity$ in allometric systems encourages us to attempt
deriving the laws (3a) and (3b) by the use of EPI. \ \ Indeed EPI has been
successfully used in a wide range of amplitude-estimation problems [3],[4],
[18]-[25] for complex systems. Its success probably traces to its basis in
Fisher information (1), which is both a measure of complexity [26], [27] and
has other important physical properties [28], [29].

All uses of the EPI principle require prior knowledge of one or more
invariances.\ Their general aim is to define the information functional $J(a)
$. \ In this problem the aim is to form a $J(a)$ that somehow represents the
$full$ $range$ of biological and physical attributes $a\equiv a_{n}$ that obey
laws (3a,b). \ $What$ $can$ $such$ $a$ $broad$ $range$ $of$ $effects$ $have$
$in$ $common?$ \ One property is that of originating on the microlevel of
biological cells or unit cells. \ Another is asymptotic behavior near the
origin. \ The following summarizes these properties:

(i) \ For all systems, information $J(a)$ originates on the $discrete$
$microlevel$. \ For example, in nonliving systems such as regular crystals or
irregular polymer chains, the sources are the unit cells or individual
molecules, respectively.\ Likewise, in a living system, biological cells are
the ultimate sources of information about a biological attribute $a$.\ 

The information $J(a)$ is assumed to propagate as a superposition of plane
waves, from a subset of cells and cell groupings to the observer. \ These
waves originate at a "unit cell" $\Delta a=1$ of $a$-space.\ \ (See
alternative (v) below.) \ The discrete nature of the "cell sources" will be
essential to the calculation. \ See Secs \ VII A,B. \ The model will also make
some useful predictions on biological $sources$ of the information (Sec. XII).

(ii) The allometric laws obey certain $asymptotic$ $behaviors$ near the
origin, as expressed next.\ \ \ \ \ \ \ \ \ \ \ \ \ \ \ \ \ \ \ \ \ \ \ \ \ \ \ \ \ \ \ \ \ \ \ \ \ \ \ \ \ \ \ \ \ \ \ \ \ \ \ \ \ \ \ \ \ \ \ \ \ \ \ \ \ \ \ \ \ 

Differentiating either allometric law (3a) or (3b)\ shows that any one
allometric law (suppressing indices $n,k$) obeys%

\begin{equation}
\frac{dy}{dx}\rightarrow\infty\text{ as }x\rightarrow0,\text{ for }a<1,\text{
but} \tag{4a}%
\end{equation}

\begin{equation}
\frac{dy}{dx}\rightarrow0\text{ as }x\rightarrow0,\text{ for }a>1. \tag{4b}%
\end{equation}
In words, the rates of increase of certain attributes increase without limit,
while others decrease without limit, as organism size $x$\ approaches zero.
\ Since the size can never equal zero (as mentioned above) the trends are
mathematically well defined. \ They also are intuitively reasonable in many
cases. Hence we make these general requirement of our solution as well.
\ Properties (4a,b) are used in Secs. IX A and E.

(iii) \ In $general$ cases (3a) of allometry: the powers $a_{n}$ are regarded
as $a$ $priori$ $fixed$ $numbers$ of unknown size (the view taken by classical
estimation theory [1]). \ These do not generally extremize (2).\ \ This
property is used in Sec. IX E.

However, in specifically $biological$ cases (3b) the $a_{n}$ are presumed to
be $optimal$ in extremizing principle (2). \ \ That is, Darwinian evolution
forces a progressive drift of organismal attributes toward those values which
confer maximal fitness on the organism. \ This model property is used in Secs.
VIII C and IX E. Maximal fitness is taken to be achieved by those attribute
values $a$ that extremize principle (2) (see Sec. XII).

(iv) \ (Only) in biological cases (3b), the independent variable $x$ is the
mass of the organism. \ That is, laws (3b) are scaling laws covering a range
of sizes, where the sizes are specified by mass values $x$. \ Why specifically
"mass," is discussed in Sec. XII. \ In nonliving systems the nature of $x$
depends upon the system. \ 

(v) (Only) in biological cases (3b). \ Alternative to the unit-cell assumption
(i) of $\Delta a=1,$ more generally allow $\Delta a=L,$ some unknown constant.
\ $L$ should be fixed by some reasonable biological requirement. \ For
example, the identification of the $a_{n}$ with pure $numbers$ requires that
one be fixed as a boundary condition. \ Then let $a_{1}\equiv1/4.$ In Sec. X
it is found that on this basis $L=1$ as before.\ 

Note that these model assumptions are not in themselves sufficient to imply
the allometric laws. \ For example, laws (3a,b) with $x$ incorrectly replaced
by $\sin(x)$ would still satisfy requirements (4a,b) of (ii).

Finally, not all systems obey allometry Eqs. (3a,b). \ Therefore, such systems
do not obey this model, by the necessity condition (3c) above.\ This is
further discussed in Sec. XII.

\section{MEASUREMENT CHANNEL FOR PROBLEM}

\ The EPI principle will be applied to both living and nonliving systems.
Thus, the measurement channel described next is, in general, that of either a
living or a nonliving system. \ However, for definiteness, biological
terminology is often used.

\textbf{A. Measurement, system function}

\ In general, the measured value $y$ of an attribute $a$ is a function%

\begin{equation}
y=Cf(x,a),~-\infty\leq a\leq+\infty,~ \tag{5}%
\end{equation}
for\ some constant $C$ and some deterministic function $f$. \ \ The latter is
called the "system" or "channel" function. \ The channel function defines how
an attribute value $y$ results from a corresponding class of attribute value
$a$ and a random source effect $x$ within the system. \ \ Here $x$ is a random
value of the mass of \ a randomly chosen system (a biological creature or a
nonliving system such as a polymer). \ The source variable $x$ obeys some
unknown and arbitrary probability law $p_{X}(x).$ \ Its details will not
matter to the calculation. \ 

The overall aim of this use of EPI will be to find the constants $C$ and the
channel functions $f(x,a)$ in the presence of any fixed but arbitrary PDF
$p_{X}(x)$ for the mass $x$. \ Hence, the functions $f(x,a)$ will be varied to
achieve the extremum that is required in Eq. (2). \ The system function will
turn out to be the allometric law (3a,b). \ In biological cases the attribute
value $a$ will be further varied to extrremize $I-J$ in Eq. (2). \ The
solution will equal $n/4$, for values of $n=0,\pm1,\pm2,...$

\ The particular $form$ of the system function $f$ defines the physics of the
particular channel. \ As a simplistic example, for some channels not
considered here, $f(x,a)=a+x.$ \ This would be the familiar case of additive
noise corrupting a signal value.

\textbf{B. Some caveats to EPI derivation}\ \ \ \ \ \ \ \ \ \ \ \ \ \ \ \ \ \ \ \ \ \ \ \ \ \ \ \ \ \ \ \ \ \ \ \ \ \ \ \ \ \ \ \ \ \ \ \ \ \ \ \ \ \ \ \ \ \ \ \ \ \ \ \ \ \ \ \ \ \ \ \ \ \ \ \ \ \ \ \ 

It should be noted that past use of EPI has been through variation of the
system $PDFs$ $or$ $amplitude$ $functions$, not through variation of their
system function $f(x,a)$\ as proposed here. \ The success of the approach in a
wide range of amplitude-estimation problems [3],[4], [18]-[22], [23], [24],
[25] implies that systems in general $obey$ $EPI$ $through$ $variation$ $of$
$their$ $PDFs$ $or$ $amplitude$ $functions$. \ However, it is not known at
this point whether systems as well $obey$ $EPI\ through$ $variation$ $of$
$their$ $channel$ $functions$. \ The derivation below will be positive in this
regard, i.e, will show that $if$ a system obeys EPI on this level, and also
the model of Sec. IV, then it obeys allometry.

In the next two sections we proceed to form the information functionals $I(a)
$ and $J(a)$, and then use them in the EPI principle (2).

\section{DATA INFORMATION $I$}

We first evaluate the information $I(a).$ \ The aim is to relate $I(a)$ to the
unknown system function $f(x,a)$, so that EPI principle (2) can be implemented
through variation of $f(x,a)$. \ \ 

From Eq. (1a), the average over $y$ explicitly gives \ \ \ \
\begin{equation}
I=I(a)=\int dy~p(y|a)\left[  \frac{\partial}{\partial a}\ln p(y|a)\right]
^{2}. \tag{6}%
\end{equation}
This takes the more specialized form (11), as shown next.

Since $x$ is random, Eq. (5) actually represents the transformation of a
random variable $x$ to a random variable $y$. \ Therefore, elementary
probability theory [30] may be used to connect the respective probability laws
$p_{X}(x)$ and $p(y|a)$, as%

\begin{equation}
p(y|a)dy=p_{X}(x)dx,~~~~\text{\ }dy>0,~dx>0. \tag{7}%
\end{equation}
We used $p_{X}(x|a)=p_{X}(x)$ since, as previously discussed, a mass value $x$
is selected independently of the choice of attribute. \ By (5), \ %

\begin{equation}
\frac{dy}{dx}=Cf~^{\prime}(x,a) \tag{8}%
\end{equation}
where the prime denotes $\partial/\partial x.$ \ Combining Eqs. (7)\ and (8) gives%

\begin{equation}
p(y|a)=\frac{p_{X}(x)}{C|f~^{\prime}(x,a)|}. \tag{9}%
\end{equation}
This is to be used in Eq. (6) to form $I$. \ \ First, taking a logarithm and
differentiating gives%

\begin{equation}
\frac{\partial}{\partial a}\ln p(y|a)=-\frac{\partial}{\partial a}%
\ln|f~^{\prime}(x,a)|. \tag{10}%
\end{equation}
Conveniently, both $p_{X}(x)$ and the constants $C$ have dropped out.\ Using
the results (7) and (10) in Eq. (6) gives%

\begin{equation}
I=\int dx~p_{X}(x)\left[  \frac{\partial}{\partial a}\ln|f~^{\prime
}(x,a)|\right]  ^{2}. \tag{11}%
\end{equation}
That is, the averaging $<~>$ is now explicitly over the random variable $x$.
\ Also, $I$ is now related to the unknown function $f(x,a)$, as was required.

\section{SOURCE INFORMATION $J(a)$\ }

\textbf{A. Microlevel contributions}

Recalling the model assumption (i) of Sec. IV, $J(a)$ originates at the cell
level. \ In general, some cells and cell groups contribute independently, and
others dependently, to $J(a)$. \ Then, by the additivity property of Fisher
information [4], the total information $J(a)$ is simply $the$ $signed$ $sum$
of $positive$ and $negative$ information contributions from the independent
cells and cell groupings of the organism. \ A well-behaved function $J(a)$ can
of course be represented over a limited $a$-interval by a $Fourier$ $series$
of such terms. \ \ What interval size should be used?

\ Here we use model assumption (i) (Sec. IV) of a unit interval. \ A unit
interval of $a$-space seems reasonable from various viewpoints. \ First, it is
fundamental to many physical effects, such as in solid state physics where the
number of degrees of freedom $per$ $unit$ $energy$ $interval$ is of
fundamental importance. \ Second, a unit interval is certainly the $simplest$
possible choice of an interval, and hence preferred on the basis of Occam's
razor. \ 

The alternative model assumption (v) (Sec. IV)\ of a $general$ interval size
$\Delta a=L$ is taken up in Sec. X.

\textbf{B. Fourier analysis}

In Sec. IV, item (i), the information $J(a)$ was modeled as propagating waves.
\ This can be substantiated. \ Heat or entropy propagates via plane
wave-Fourier series [31],[32]. Fisher information $J(a,t)$ is, like entropy, a
measure of disorder, monotonically decreasing with an increase in time $t$
[3],[4],[20]. \ Moreover, both the flux of heat/disorder [32] and the flow of
information $J(a,t)$ obey Fokker-Planck equations. \ We assume steady-state
boundary conditions so that $J(a,t)=J(a)$. \ (The attributes supply
information at a constant rate in time.) \ The general solution of this
Fokker-Planck equation over a unit interval of $a$ \ (as above) is a simple
Fourier series [31],[32]%

\begin{align}
\underset{0\leq a\leq1}{J(a)}  &  =%
%TCIMACRO{\dsum \limits_{m}}%
%BeginExpansion
{\displaystyle\sum\limits_{m}}
%EndExpansion
F_{m}\exp\left(  2\pi ima\right)  ,\ ~~F_{m}=%
%TCIMACRO{\dint \limits_{0}^{1}}%
%BeginExpansion
{\displaystyle\int\limits_{0}^{1}}
%EndExpansion
da^{\prime}J(a^{\prime})\exp(-2\pi ima^{\prime})\tag{12a}\\
~~\ J(a)  &  \geq0,~~i=\sqrt{-1}.\nonumber
\end{align}

However, this series is inadequate for our purposes. \ First, Eqs. (3a,b) hold
over an infinite range $-\infty\leq a\leq\infty$ of attribute values, not only
over a unit interval. \ Second, we expect function $J(a)$ to be an even function,%

\begin{equation}
J(a)=J(-a) \tag{12b}%
\end{equation}
since there is no reason to expect a negative attribute value to provide more
information than its corresponding positive value.\ \ One way to accomplish
the range $-\infty\leq a\leq\infty$ is to form the Fourier series for $J(a)$
over $a$ $sequence$ of symmetrically placed, half-unit interval pairs
$(-1/2\leq a\leq0)$ and $(0\leq a\leq1/2);$ $(-1\leq a\leq-1/2)$ and $(1/2\leq
a\leq1);$ etc. \ These are denoted as%

\begin{equation}
a=\pm\left(  \frac{j}{2},\frac{j+1}{2}\right)  ,~j=0,1,2,.... \tag{12c}%
\end{equation}
Each interval number $j$\ defines in this way a total $unit$ interval for $a$,
as required. \ The $half$-unit intervals (12c) are contiguous and span all of
$a$-space. \ The $J(a)$ for each interval obeys [31]%

\begin{align}
\underset{\pm(j/2,(j+1)/2)}{J(a)}  &  =%
%TCIMACRO{\dsum \limits_{m}}%
%BeginExpansion
{\displaystyle\sum\limits_{m}}
%EndExpansion
B_{mj}\exp\left(  4\pi ima\right)  ,\ ~~J(a)\geq0\tag{12d}\\
B_{mj}  &  =2%
%TCIMACRO{\dint \limits_{j/2}^{(j+1)/2}}%
%BeginExpansion
{\displaystyle\int\limits_{j/2}^{(j+1)/2}}
%EndExpansion
da^{\prime}J(a^{\prime})\exp(-4\pi ima^{\prime}),~~j=0,1,2,...\nonumber
\end{align}
Thus each value of $j$ identifies an interval over which $J(a)$ is defined by
a distinct set of Fourier coefficients $B_{mj},$ $m=1,2,...$ Since these
intervals (12c) are contiguous and span $a$-space, the resulting $J(a)$ is
defined over all $a$-space as required. \ The factors $4$ in the exponents,
which will prove decisive, arise because each $a^{\prime}$-integration (12d)
is over an interval of length $1/2$ (rather than $1$ as in (12a)).

This simplifies further. \ Because each $J(a)$ is an information and therefore
$real$, (12d) becomes%

\begin{equation}
\underset{\pm(j/2,(j+1)/2)}{J(a)}=%
%TCIMACRO{\dsum \limits_{m}}%
%BeginExpansion
{\displaystyle\sum\limits_{m}}
%EndExpansion
B_{mj}^{(re)}\cos\left(  4\pi ma\right)  -%
%TCIMACRO{\dsum \limits_{m}}%
%BeginExpansion
{\displaystyle\sum\limits_{m}}
%EndExpansion
B_{mj}^{(im)}\sin\left(  4\pi ma\right)  ,~\ ~j=0,1,2,... \tag{12e}%
\end{equation}
where $(re)$ and $(im)$ denote real and imaginary parts. \ 

Requirement (12b) of symmetry can only be obeyed if generally $B_{mj}%
^{(im)}=0$ for all $m,$ so that%

\begin{equation}
\underset{\pm(j/2,(j+1)/2)~}{J(a)}=%
%TCIMACRO{\dsum \limits_{m}}%
%BeginExpansion
{\displaystyle\sum\limits_{m}}
%EndExpansion
A_{mj}\cos\left(  4\pi ma\right)  ,~~~A_{mj}\equiv B_{mj}^{(re)},~~j=0,1,2,...
\tag{12f}%
\end{equation}
Next, using $B_{mj}^{(im)}=0$ and that $J(a^{\prime})$ is real in the 2nd Eq.
(12d) indicates that%

\begin{equation}
B_{mj}=2%
%TCIMACRO{\dint \limits_{j/2}^{(j+1)/2}}%
%BeginExpansion
{\displaystyle\int\limits_{j/2}^{(j+1)/2}}
%EndExpansion
da^{\prime}~J(a^{\prime})\cos4\pi ma^{\prime}=B_{mj}^{(re)}\equiv
A_{mj},~~j=0,1,2... \tag{12g}%
\end{equation}
By Eq. (2), $J(a)$\ must obey positivity [3],[4]. \ Therefore, the
coefficients $A_{mj}$ must be constrained to give positive or zero values
$J(a)$ at all $a.$

\section{PARTICULAR EPI PROBLEM}

For generality of results, in the analysis that follows we will regard the
cellular contributions $A_{mj}$ in (12f) as arbitrary, except for causing
symmetry (12b) and positivity (12d) in $J(a)$.

Using the particular informations (11) and (12f) in the general EPI principle
(2) gives a problem%

\begin{align}
I-J~~\ \ ~  &  =~\int dx~p_{X}(x)\left[  \frac{\partial}{\partial a}%
\ln|f~^{\prime}(x,a)|\right]  ^{2}~~~~\ \ \ \ \tag{13}\\
-\int dx~p_{X}(x)%
%TCIMACRO{\dsum \limits_{m}}%
%BeginExpansion
{\displaystyle\sum\limits_{m}}
%EndExpansion
A_{mj}\cos\left(  4\pi ma\right)   &  =~extremum,~~~\ \ j=0,1,2,...\nonumber
\end{align}
Here a choice of $a$ defines 1:1 a choice of interval $j$, via Eq. (12c), and
therefore a choice of coefficients $A_{mj},$ $m=1,2,...$ \ For mathematical
convenience, we appended a multiplier of $1$ (a normalization integral $\int
dx~p_{X}(x)$) to the second sum $J$.

As discussed in Sec. V\ A, we seek the channel functions $f~(x,a)$ and (in
biological cases) the system parameters $a$ that extremize (13), in the
presence of any $fixed$ source PDF $p_{X}(x).$ \ Accordingly, the extremum in
the principle (13) is first attained through variation of functions $f(x,a)$
and then, in biological cases, through the additional variation of parameters
$a$. \ The mass PDF $p_{X}(x)$ is $not$ varied, and turns out to not affect
the answer. \ Thus, $the$ $channel$ $is$ $optimized$ $in$ $the$ $presence$
$of$ $a$ $given$ $source.$

\textbf{A. Synopsis of the approach}

\ \ The basic approach consists of three overall steps, as carried through in
Sec. VIII B - Sec. IX E: \ 

(1) The information flow $I-J$ is extremized through choice of system function
$f(x,a)$, in the presence of any fixed PDF\ mass law $p_{X}(x).$ \ This gives
a general power law for its derivative $\partial f(x,a)/\partial x\equiv
f~^{\prime}(x,a),$%

\begin{equation}
f~^{\prime}(x,a)=h(x)^{a-1},~a\text{ real.} \tag{14a}%
\end{equation}
[Eq. (20)]. \ Quantity $h(x)$ is\ some unknown base function of $x.$ \ 

(2) The base function $h(x)$ is found, by further extremizing $I-J$ with
respect to it, giving $h(x)=b_{1}x$ [Eq. (38)]. \ Using this in (14a) gives%

\begin{equation}
f(x,a)=x^{a} \tag{14b}%
\end{equation}
[Eq. (42)] after an integration. \ An irrelevent constant is ignored. By Eq.
(5), this achieves derivation of the general allometric law (3a). \ \ 

(3)\ \ Finally, for a system that is biological, $I-J$ is extremized with
respect to the choice of $a$, which gives $a=n/4$ [Eq. (25)]. \ Using this in
(14b) gives%

\begin{equation}
f(x,a)=x^{n/4}. \tag{14c}%
\end{equation}
This is the biological allometric law (3b). \ The approach (1)-(3) is now
carried through.

\textbf{B. Primary variation of the system function leads to a family of
power-laws}

The aim is to find the channel function $f(x,a)$ in the presence of a fixed
source function $p_{X}(x).$\ \ Hence we first vary $f(x,a)$, by use of the
calculus of variations, holding the function $p_{X}(x)$ constant.
\ Conveniently, it will drop out during the variation. \ The Lagrangian for
the problem is, by definition, the integrand of (13)%

\begin{equation}%
%TCIMACRO{\tciLaplace}%
%BeginExpansion
\mathcal{L}%
%EndExpansion
=~p_{X}(x)\left[  \frac{\partial}{\partial a}\ln g(x,a)\right]  ^{2}-p_{X}(x)%
%TCIMACRO{\dsum \limits_{m}}%
%BeginExpansion
{\displaystyle\sum\limits_{m}}
%EndExpansion
A_{mj}\cos\left(  4\pi ma\right)  ,~~j=0,1,2,... \tag{15a}%
\end{equation}
where we introduced a new function $g$ defined as%

\begin{equation}
|f~^{\prime}(x,a)|~\equiv g(x,a). \tag{15b}%
\end{equation}
In this way the function $g(x,a)$ replaces $f~(x,a)$ as the quantity to vary
in (15a). \ Keeping in mind that the PDF $p_{X}(x)$ on mass remains a $fixed$
function during the variation, the Lagrangian (15a) is readily differentiated as%

\begin{equation}
\frac{\partial%
%TCIMACRO{\tciLaplace}%
%BeginExpansion
\mathcal{L}%
%EndExpansion
}{\partial(\partial g/\partial a)}=2~p_{X}(x)\frac{\partial g/\partial
a}{g^{2}}\text{ \ and}~\ \ \frac{\partial%
%TCIMACRO{\tciLaplace}%
%BeginExpansion
\mathcal{L}%
%EndExpansion
}{\partial g}=-2~p_{X}(x)\frac{\left(  \partial g/\partial a\right)  ^{2}%
}{g^{3}},~~\ g\equiv g(x,a). \tag{16}%
\end{equation}
Using these in the Euler-Lagrange equation [31]%

\begin{equation}
\frac{\partial}{\partial a}\left(  \frac{\partial%
%TCIMACRO{\tciLaplace}%
%BeginExpansion
\mathcal{L}%
%EndExpansion
}{\partial(\partial g/\partial a)}\right)  =\frac{\partial%
%TCIMACRO{\tciLaplace}%
%BeginExpansion
\mathcal{L}%
%EndExpansion
}{\partial g} \tag{17}%
\end{equation}
gives, after some trivial cancellation,%

\begin{equation}
\frac{\partial}{\partial a}\left[  \frac{\partial g/\partial a}{g^{2}}\right]
=-\frac{\left(  \partial g/\partial a\right)  ^{2}}{g^{3}},~\ \ ~g\equiv
g(x,a). \tag{18}%
\end{equation}
Thus, the unknown PDF $p_{X}(x)$ has dropped out, as we anticipated above.
\ \ Doing the indicated differentiation gives after some algebra%

\begin{equation}
g\frac{\partial^{2}g}{\partial a^{2}}-\left(  \frac{\partial g}{\partial
a}\right)  ^{2}=0. \tag{19}%
\end{equation}
\ \ The general solution to this can be found by using $g\equiv\exp(k),$
$k\equiv k(x,a),$ in (19) and solving the resulting differential equation for
$k$. \ The answer is $k=K(x)a+L(x)$, with $K(x),L(x)$ arbitrary functions.
\ Exponentiating back to $g$ gives an answer%

\begin{equation}
g(x,a)=h(x)^{a-1}, \tag{20}%
\end{equation}
where $h(x)\equiv\exp(K(x))$ is an arbitrary real function of $x$ called the
"base function," \ and we took $L(x)\equiv-K(x).$ The latter choice gives the
term $-1$ in the exponent of (20), for later numbering of the attributes (see
also (v), Sec. IV). \ The solution (20) may be readily shown to satisfy
differential equation (19), keeping in mind that its derivatives are with
respect to $a$ and not $x$.

Hence the solution to the problem has the general form of a $power$ $law$.
\ That is, on the basis of optimal information flow $J\rightarrow I$, nature
generally acts to form power-law solutions for the rate of change $g(x,a)$\ of
the channel function.

The general solution (20) contains a general base function $h(x)$ of the mass.
This function will be found in Sec. IX. \ \ Also, the values of the power
$(a-1)$ of $h(x)$ to be used for the biological laws are not yet fixed.
\ These unknown powers will next be fixed, as the 2nd optimization step.

\textbf{C. Variation of the attribute parameters gives powers }%
$\mathbf{a\equiv a}_{n}=\mathbf{n/4}$

Here, by premise (iii) of Sec. IV, we vary $a$, for use in the biological
laws. \ (Note that this will not affect the general law (3a) derivation since
$a$ so obtained [Eq. (25)] will $not$ be used in that derivation.) \ Since $a$
is a discrete variable, ordinary calculus is used, differentiating
$\partial/\partial a$ Eq. (13) and equating the result to zero. \ This gives,
after use of (15b),%

\begin{align}
&  \frac{\partial}{\partial a}\int dx~p_{X}(x)\left[  \frac{\partial
g(x,a)/\partial a}{g(x,a)}\right]  ^{2}\tag{21}\\
&  -\frac{\partial}{\partial a}\left[  \int dx~p_{X}(x)%
%TCIMACRO{\dsum \limits_{m}}%
%BeginExpansion
{\displaystyle\sum\limits_{m}}
%EndExpansion
A_{mj}\cos\left(  4\pi ma\right)  \right] \nonumber\\
&  =0,~~~~~j=0,1,2...\nonumber
\end{align}

\ The first derivative term in (21) is next shown to be zero. \ Its derivative
$\partial/\partial a$ operation may be moved to within the integrand, giving%

\begin{equation}
\frac{\partial}{\partial a}\left(  \frac{g_{a}}{g}\right)  ^{2},\text{
\ \ \ \ }g_{a}\equiv\frac{\partial g(x,a)}{\partial a}. \tag{22}%
\end{equation}
Carrying out the indicated derivative $\partial/\partial a$ gives%

\begin{equation}
2\left(  \frac{g_{a}}{g}\right)  \frac{\partial}{\partial a}\left(
\frac{g_{a}}{g}\right)  =2\left(  \frac{g_{a}}{g}\right)  \left(
\frac{gg_{aa}-g_{a}^{2}}{g^{2}}\right)  =0 \tag{23}%
\end{equation}
by Eq. (19). \ Eq. (19) could be used since the biological optimization
requires the $simultaneous$ satisfaction of both conditions (17) and (21).

We showed in the preceding paragraph that the the left-hand term in (21)
becomes zero after the indicated differentiation, that is, $\partial
I/\partial a=0.$ This has two important consequences. \ First, as will be
shown below, $I$ then does not depend upon $a$ for the power law solution
(20). \ 

Second, only the right-hand term of (21) now remains. \ It defines a problem%

\begin{align}
&  \frac{\partial}{\partial a}\left[  p_{X}(x)%
%TCIMACRO{\dsum \limits_{m}}%
%BeginExpansion
{\displaystyle\sum\limits_{m}}
%EndExpansion
A_{mj}\cos\left(  4\pi ma\right)  \right] \tag{24}\\
&  =-p_{X}(x)%
%TCIMACRO{\dsum \limits_{m}}%
%BeginExpansion
{\displaystyle\sum\limits_{m}}
%EndExpansion
A_{mj}\left(  4\pi m\right)  \sin\left(  4\pi ma\right)
=0,~~~~j=0,1,2,...\nonumber
\end{align}
(Note that $\partial\cos(4\pi ma)/\partial a=-4\pi m\sin(4\pi ma)$ within any
interval $j$.) \ For arbitrary coefficients $A_{mj},$ the required zero is
obtained if and only if%

\begin{equation}
a\equiv a_{n}=\frac{n}{4},~~n=0,\pm1,\pm2,\pm3,... \tag{25}%
\end{equation}
since then the sine function in (24) becomes $\sin(mn\pi)=0$ for all integers
$m,n$. \ Note that the solution values (25) form $in$ $sequence$ for the
different unit intervals $j$ given by (12c). \ As examples: The interval for
$j=0$ is $(-1/2,0)$, $(0,1/2)$ and contains solution values (25)
$a=0,\pm1/4,\pm2/4$. \ The interval for $j=1$ is $(-1,-1/2)$, $(1/2,1)$ and
contains solutions $a=\pm2/4,\pm3/4,\pm4/4$. \ And so on, thereby forming
$all$ solutions (25).

Result (25) shows that the attribute value $a$ must be a multiple of $1/4$,
or, $the$ $powers$ $in$ $the$ $power$ $law$ (20) $are$ $multiples$ $of$ $1/4$.
\ This is an important milestone in the biological derivation. \ We emphasize
that it only could follow because of the $discrete$ nature of the sum over
$m,$ which follows from the model assumption (i) (Sec. IV) that information
originates on the level of the discrete cells.\ 

\section{SECONDARY EXTREMIZATION THRU CHOICE OF $h(x)$}

\ The solution (20) to the extremization problem (13) of $I-J=extrem$. was
found to contain an arbitrary function $h(x)$ . \ Clearly the appropriate
$h(x)$ is the one that further extremizes $I-J$. \ We seek this function here.
\ First we establish a general property of $h(x).$

\textbf{A. \ Special form of function }$\mathbf{h(x)}$\textbf{\ }

Here we show that $h(x)$ can be expressed as a linear term in $x$ plus a
function that is at least quadratic in $x$. \ Function $h(x)$ can be generally
expanded in Taylor series as%

\begin{equation}
h(x)=b_{0}+b_{1}x+b_{2}x^{2}+b_{3}x^{3}+... \tag{26}%
\end{equation}
Differentiating (5), then using (26) in (15b) and (20) gives, in sequence,%

\begin{align}
\frac{dy_{nk}}{dx}  &  =C_{nk}\frac{df(x,a_{n})}{dx}=C_{nk}g(x,a_{n}%
)=C_{nk}h(x)^{a_{n}-1}\tag{27}\\
&  =C_{nk}(b_{0}+b_{1}x+b_{2}x^{2}+...\ )^{a_{n}-1}.\text{ \ }\nonumber
\end{align}
Then%

\begin{equation}
\lim_{x\rightarrow0}\frac{dy_{nk}}{dx}=C_{nk}b_{0}^{a_{n}-1}\equiv\frac
{C_{nk}}{b_{0}^{1-a_{n}}}. \tag{28}%
\end{equation}

We now use the model properties (4a), (4b). \ If $a_{n}<1$ then limit (4a)
holds. \ This can only be obeyed by (28)\ if%

\begin{equation}
b_{0}=0. \tag{29}%
\end{equation}
Consequently, by (26) $h(x)=b_{1}x+b_{2}x^{2}+b_{3}x^{3}+...$ or%

\begin{equation}
\text{ }h(x)=b_{1}x+[k(x)]^{2},~~k(x)\equiv x\sqrt{b_{2}+b_{3}x+...} \tag{30}%
\end{equation}
for some function $k(x)$. \ \ By the square root operation in\ (30), the
latter is in general either pure real or pure imaginary at each $x$; it is
found next.

\textbf{B. Resulting variational principle in base function }$\mathbf{h(x)}$

Using definition (15b), and Eq. (20) in Eq. (11), gives an information level%

\begin{align}
I  &  =\left\langle \left[  \frac{\partial}{\partial a}\ln\left(
h(x)^{a-1}\right)  \right]  ^{2}\right\rangle =\left\langle \left[
\frac{\partial}{\partial a}(a-1)\ln h(x)\right]  ^{2}\right\rangle \tag{31}\\
&  =\left\langle \ln^{2}h(x)\right\rangle \nonumber
\end{align}
after obvious algebra. \ Quantity $a$ has dropped out.

The information difference $I-J$ is to be extremized in a $total$ $sense$.
\ The base function $h(x)$ that defines $I$ in (31) has been expressed in
terms of a new function $k(x)$ [Eq. (30)]. \ Hence $I-J$ must be further
(secondarily) extremized $through$ $variation$ $of$ $function$ $k(x)$.
\ \ Using EPI result (30) in (31), \ and combining this with (12f)\ and (25),
gives a new problem%

\begin{equation}
I-J=\left\langle \ln^{2}\left[  b_{1}x+k^{2}(x)\right]  \right\rangle -%
%TCIMACRO{\dsum \limits_{m}}%
%BeginExpansion
{\displaystyle\sum\limits_{m}}
%EndExpansion
A_{mj}(-1)^{mn}\equiv extremum \tag{32}%
\end{equation}
in $k(x)$.

\textbf{C. Secondary variational principle in associated function
}$\mathbf{k(x)}$

Since the $A_{m}$ are independent of $k(x)$, the net Lagrangian in (32) for
varying $k(x)$\ is%

\begin{equation}%
%TCIMACRO{\tciLaplace}%
%BeginExpansion
\mathcal{L}%
%EndExpansion
=~p_{X}(x)\ln^{2}\left[  b_{1}x+k^{2}(x)\right]  . \tag{33}%
\end{equation}
Function $p_{X}(x)$ arises out of the expectation operation $<\ >$\ in (32),
and is also independent of $k(x)$. \ The general Euler-Lagrange equation for
problem (33) is [31]%

\begin{equation}
\frac{d}{dx}\left(  \frac{\partial%
%TCIMACRO{\tciLaplace}%
%BeginExpansion
\mathcal{L}%
%EndExpansion
}{\partial k^{\prime}(x)}\right)  =\frac{\partial%
%TCIMACRO{\tciLaplace}%
%BeginExpansion
\mathcal{L}%
%EndExpansion
}{\partial k(x)},~~\ ~\ k^{\prime}(x)\equiv dk/dx. \tag{34}%
\end{equation}
Since $%
%TCIMACRO{\tciLaplace}%
%BeginExpansion
\mathcal{L}%
%EndExpansion
$ in (33) does not depend upon $k^{\prime}(x),$ the left-hand side of (34) is
zero. \ Also, differentiating (33) gives%

\begin{equation}
\frac{\partial%
%TCIMACRO{\tciLaplace}%
%BeginExpansion
\mathcal{L}%
%EndExpansion
}{\partial k(x)}=\frac{2~p_{X}(x)\ln\left[  b_{1}x+k^{2}(x)\right]  }%
{b_{1}x+k^{2}(x)}2k(x)\equiv0. \tag{35}%
\end{equation}
Once again, $p_{X}(x)$ is merely a constant multiplier, dropping out of the
problem.\ Eq. (35) has two formal solutions.

\textbf{D. Result }$\mathbf{k(x)=0,}$ \textbf{giving base function
}$\mathbf{h(x)}$\textbf{\ proportional to }$\mathbf{x}$

The first formal solution to (35) is%

\begin{equation}
b_{1}x+k^{2}(x)\equiv h(x)=1, \tag{36}%
\end{equation}
the middle identity by (30). \ The second solution is%

\begin{equation}
k(x)=0. \tag{37}%
\end{equation}
(Notice that this holds regardless of whether $k(x)$ is pure real or pure imaginary.)

However, one solution is readily eliminated. \ The candidate (36) when used in
(31) gives $I=\left\langle \left[  \ln1\right]  ^{2}\right\rangle =0.$ \ This
extremum is the $absolute$ $minimum$ value possible for Fisher information.
\ However, $I=0$ is rejected since then the observed value $y$ of the
attribute would unrealistically provide no information about the attribute.
\ \ Hence the solution (36) is rejected. \ 

By comparison, the candidate (37) when used in (30) gives
\begin{equation}
h(x)=b_{1}x,\ \tag{38}%
\end{equation}
and consequently%

\begin{equation}
I\equiv I_{extrem}=\left\langle \left[  \ln(b_{1}x)\right]  ^{2}\right\rangle
. \tag{39}%
\end{equation}
by (31). \ Information (39) is generally nonzero, thereby representing a
subsidiary minimum, which\ makes sense on the grounds that the observation
must contain at least some information. \ Hence the solution (37), (38) is accepted.\ 

\textbf{E. Final allometric laws}\ 

We are now in a position to form the final allometric laws (3a,b) for,
respectively, general and living systems. \ Substituting the solution (38)
into Eqs. (15b) and (20) gives $|f~^{\prime}(x,a)|~=(b_{1}x)^{a-1}$ or%

\begin{equation}
f~^{\prime}(x,a)|~=\pm(b_{1}x)^{a-1}. \tag{40}%
\end{equation}
Indefinitely integrating gives%

\begin{equation}
f(x,a)=\pm b_{1}^{a-1}\int dxx^{a-1}\equiv\pm a\int dxx^{a-1}, \tag{41}%
\end{equation}
for a suitably defined $b_{1}.$\ An additive constant in (41) is taken to be
zero by asymptotic prior knowledge (4b): \ In all attribute parameter cases
$a>1,$ as $x\rightarrow0$ it is required that the attribute value
$y_{nk}\rightarrow0,$ and hence by Eq. (5) likewise $f(x,a)\rightarrow0.$ The
integral (41) is directly evaluated as%

\begin{equation}
f(x,a)=x^{a}. \tag{42}%
\end{equation}
We used the fact that the attribute values $y$ are positive [Eqs. (3a,b)] in
order to rule out the negative alternative.

The $general$ allometric law (3a) is to hold for a priori empirically defined
powers $a_{n}$ (see (iii), Sec. IV).\ Here the specific powers (25) that held
for optimization of $I-J$ do not apply. \ The solution is more simply the
combination of Eqs. (42)\ and (5). \ Reinserting subscripts gives%

\begin{equation}
y_{nk}\equiv C_{nk}f~(x,a_{n})=Y_{nk}x^{a_{n}},~~~\text{so that}%
~\ \ ~Y_{nk}\equiv C_{nk}\ ,~~~~\ n=0,\pm1,\pm2,...~ \tag{43}%
\end{equation}
This confirms the general allometric law (3a) for empirically known $a_{n}.$

Next we turn to the $biological$ allometric law, which is modelled ( (iii),
Sec. IV) to hold for the particular powers $a_{n}$ given by (25) that enforce
a further extremization in the problem (2). \ Using powers (25) in the power
law solution (42), and also using (5), gives%

\begin{equation}
y_{nk}\equiv C_{nk}f(x,a_{n})=Y_{nk}x^{n/4},~\text{so that}~~~Y_{nk}\equiv
C_{nk},~~~~\ n=0,\pm1,\pm2,... \tag{44}%
\end{equation}
This is the law (3b). \ As contrasted with laws (43), the powers $a_{n}$\ are
here purely multiples of $1/4.$ \ 

\section{ALTERNATIVE MODEL $\Delta a=L$}

The preceding derivation assumed a priori a unit fundamental length $\Delta
a=1$ ((i), Sec. IV). \ \ A stronger derivation would allow $\Delta a=L$ with
$L$ general. \ With $\Delta a=L,$\ the half-unit interval pairs in Sec. VII B
are replaced with pairs of length $L/2$. \ Also, Eqs. (12c)-(12g) now hold
[31] under the replacements $j\rightarrow jL,$ $(j+1)\rightarrow(j+1)L,$ and
$m\rightarrow m/L.$ \ Consequently the requirement of zero for Eq. (24) now
becomes one of zero for $\sin(4\pi ma/L).$ \ The solution is $a\equiv
a_{n}=nL/4.$ Hence, by (43)\ the biological power law is now $y_{nk}%
=Y_{nk}x^{nL/4}$ instead of (44). \ \ Also, now $a_{1}=1\cdot L/4=L/4$. \ But
by model assumption (v) of Sec. IV, $a_{1}\equiv1/4.$ \ It results that $L=1$.
\ Consequently the quarter-power law (44) results once again. \ 

\section{SUMMARY}

After introducing Fisher data information $I$ in Sec. I, the information is
used in the EPI principle (2) of Sec. II. \ The general allometric laws of
science are discussed in Sec. III A, a subset from biology is discussed in
Sec. III B, and past explanations of biological allometry are discussed in
Sec. III C. \ The limited scope of the EPI derivation is discussed in Sec. III
C. \ The prior knowledge assumed in the EPI derivation is given in Sec. IV.
\ This includes the assumption that the source information $J$ originates at
the level of discrete cells, and propagates from there into measurement space.
\ Also used is specific limiting behavior of the allometric laws near the
origin. \ Caveats to the approach are discussed in Sec. V B and below in Sec.
XII. \ The rest of the paper is concerned with deriving the allometric laws
from these assumptions. \ The derivation concurrently applies to both
inanimate and biological cases. \ A brief synopsis of the mathematics of the
approach is given in Sec. VIII A.

The detailed approach is given in Secs. VIIIB\ \ - IX E, with an alternative
aspect addressed in Sec. X.

\section{DISCUSSION}

This paper has the limited aim (3c) of $establishing$ $necessity$ for
allometry. \ It shows that if a system obeys the model of Sec. IV and also
obeys EPI through variation of its channel function $f(x,a)$, it must obey
allometry. \ However, this does not necessarily imply the converse -- that any
system that obeys allometry must also obey EPI and the model. \ (Note that
this in fact might be true, but is regarded as outside the scope of the
paper.) \ Also, $not$ $all$ systems obey allometry. \ Then, by the necessity
(3c) proven in this paper, such systems do not obey the model of Sec.\ IV
and/or EPI.

By the overall approach, the allometric laws (3a,b) follow as the effect of a
flow of information $J\rightarrow I$ from an attribute source to an observer.
\ We saw that the derivation for general laws (3a) slightly differs from that
for biological laws (3b). \ Each general law (3a) accomplishes an
extremization of the loss of information $I-J$ through variation of the system
function $f(x,a_{n})$ and its subfunctions $h(x)$ and$\ k(x).$ \ By
comparison, the biological allometric laws (3b) accomplish the extremization
with respect to both these functions $and$ the system parameters $a_{n}.$ The
extra optimization with respect to the $a_{n}$ reflects the specialized nature
of biological allometry. \ But, why should biological systems be so specialized?

The answer is that, as compared to nonliving systems, biological systems have
resulted from Darwinian $evolution$. \ Thus, evolution is postulated ((iii),
Sec. IV) as selecting particular attribute parameters $a_{n}$ that\ optimize
the information flow loss $I-J$.\ \ The postulate is reasonable. \ Survival
and proliferation within an adaptive landscape favors optimization of
phenotypic traits which, in turn, confers maximal fitness on the individual.
\ Here the phenotype traits are, in fact, the attribute parameters $a_{n}%
.$\ \ Therefore, the $a_{n}$ will evolve into those values that favor maximal
fitness. \ Meanwhile,\ maximal fitness has been shown [4], [23] to result from
optimal information flow loss $I-J=extrem$. \ (The latter gives rise to the
Lotka-Volterra equations of growth which, in turn, imply maximal fitness
through "Fisher's theorem of genetic change.") \ Therefore, it is reasonable
that the same parameter values $a_{n}$ that satisfy evolution will also
satisfy $I-J=extrem.$

In a related derivation [24], under the premise that $in$ $situ$ cancer is
likewise in an evolutionary extremized state -- now of transmitting $minimal$
information about its age and size -- the EPI output result is the correct law
of cancer growth, again a power law form (3a). \ However, here $x$ is the time
and $a_{n}=1.618...$ is the Fibonacci golden mean. \ Also, as here, the
information is optimized with respect to the exponent $a_{n}.$ \ This is also
further evidence that the premise ((iii), Sec. IV) of evolutionary efficiency
is correct.

It was assumed as prior knowledge ((iv), Sec. IV) that in biological cases
(3b) the independent variable $x$ is the $mass$ of the organism. \ That is,
laws (3b) are scaling laws covering a range of sizes, where the sizes are
specified uniquely by mass values $x$. \ Aside from being a postulate of the
derivation, this is reasonable on evolutionary grounds. \ By its nature, the
process of evolution favors systems that are close to being $optimized$ with
respect to the energy [and information] they distribute [9] to phenotypic
traits at its various scales. On this basis only a dependence upon absolute
size or mass would remain.

The precise form of the biological function $J(a)$\ is unknown. \ It is
possible that it is periodic, repeating itself over each fundamental interval
$j$. \ This implies that all $A_{mj}=A_{m},$ $m=1,2,...$ irrespective of $j$.
\ Interestingly, for such periodicity $J(a)$ breaks naturally into 2 classes.
\ Back substituting any one coefficient (25) into the Fourier representation
(12f) now gives%

\begin{equation}
J(a_{n})=J(n/4)=%
%TCIMACRO{\dsum \limits_{m}}%
%BeginExpansion
{\displaystyle\sum\limits_{m}}
%EndExpansion
A_{m}\cos(mn\pi)=%
%TCIMACRO{\dsum \limits_{m}}%
%BeginExpansion
{\displaystyle\sum\limits_{m}}
%EndExpansion
A_{m}(-1)^{mn}. \tag{45}%
\end{equation}
Since the $A_{m}$ remain $arbitrary$, this still represents an arbitrary
information quantity $J(a_{n})$ for $n=0$ or $1$. \ However, for higher values
of $n$ the form (45) repeats, giving
\begin{equation}
J(\pm a_{3})=J(\pm a_{5})=...=J(\pm a_{1}),\text{ \ } \tag{46a}%
\end{equation}
and%

\begin{equation}
\text{\ }J(\pm a_{2})=J(\pm a_{4})=...=J(\pm a_{0}). \tag{46b}%
\end{equation}
Hence the odd numbered attributes $n=\pm1,\pm3,\pm5,...$ all share one fixed
level of ground truth information $J$ about their values $a_{n}$, and the even
numbered attributes $n=0,\pm2,\pm4,\pm6,...$ share another. \ Consequently,
the source information of the channel is specified by but $two$ independent
values, (say) $J(a_{0})$ and $J(a_{1}).$ \ Or, the allometric relations result
from two basic sources of information.\ As we found, the numerical values of
the two information levels remain arbitrary, since the coefficients $A_{m}$
are arbitrary. \ 

\ Finally, it is worthwhile considering why $biologically$ there should be but
two classes of information.\ The postulate (i) of Sec. IV that discrete cells
are the sources of information enters in once again. \ \ This ultimately gave
rise to the sum (12f) representing the source information $J(a)$ for the
attribute. \ The sum\ is over\ the biological cells and, by Eqs. (46a,b) there
are only two independent information sources. \ \ On this basis each cell must
provide $two$ independent sources of attribute information. \ The existence of
two such sources is, in fact, consistent with recent work [33] which concludes
that $cellular$ $DNA$ and $cellular$ $transmembrane$ $ion$ $gradients$ are the sources.

\begin{acknowledgments}
We thank Prof. Daniel Stein of the Courant Inst., N.Y.U., for checking the
math and making valuable suggestions on the modeling. \ Prof. Patrick McQuire
of the Center for Astrobiology (CSIC/INTA), Madrid, provided valuable comments
on biological aspects of the manuscript. \ Finally, valuable support was
provided by H. Cabezas of the Sustainable Technologies Division, Environmental
Protection Agency.
\end{acknowledgments}

\section*{REFERENCES}

[1] R.A. Fisher, $Statistical$ $Methods$ $and$ $Scientific$ $Inference,$ $2nd$
$ed.$ (Oliver \& Boyd, London, 1959).

[2] A. Stuart and J.K. Ord, $Kendall\prime s$ $Advanced$ $Theory$ $of$
$Statistics,$ $5th$ $ed.$, vol. 2 (Edward Arnold, London, 1991).

[3] B.R. Frieden and B.H. Soffer, Phys. Rev. E \textbf{52}, 2274 (1995)

[4] B.R. Frieden, $Physics$ $from$ $Fisher$ $Information$ (Cambridge Univ.
Press, 1998); $Science$ $from$ $Fisher$ $Information,~2nd~ed.$ (Cambridge
Univ. Press, 2004)

[5] $Scale$ $Invariance,$ $Interfaces$ $and$ $Non-Equilibrium$ $Dynamics$,
eds. A. McKane, M. Droz, J. Vannimenus and D. Wolf (Plenum Press, N.Y., 1995)

[6] W.J. Jungers, $Size$ $and$ $Scaling$ $in$ $Primate$ $Biology$ (Plenum
Press, N.Y., 1985)

[7] T. Nishikawa, A.E. Motter, Y-C Lai and F.C. Hoppensteadt, Phys. Rev. Lett.
\textbf{91}, 014101 (2003)

[8] G.B. West, J.H. Brown, B.J. Enquist, Science \textbf{276}, 122 (1997);
B.J. Enquist and K.J. Niklas, Nature \textbf{410}, 655 (2001). \ 

[9] See the highly interesting article "Life's Universal Scaling Laws" by G.B.
West and J.H. Brown in Physics Today \textbf{57} (9), 36 (2004), and its references.

[10] T.A. McMahon and J.T. Bonner, $On$ $Size$ $and$ $Life$ (Scientific
American Library, N.Y., 1983)

[11] K. Schmidt-Nielsen, $Scaling:Why$ $is$ $Animal$ $Size$ $so$ $Important?$
(Cambridge University Press, Cambridge, U.K., 1984)

[12] W.A. Calder, $Size,$ $Function$ $and$ $Life$ $History$ (Harvard
University Press, Cambridge, Massachusetts, 1984)

[13] M.J. Reiss, $The$ $Allometry$ $of$ $Growth$ $and$ $Reproduction$
(Cambridge University Press, Cambridge, U.K., 1989)

[14] J.T. Bonner, $Size$ $and$ $Cycle$ (Princeton University Press, N.J., 1965)

[15] W.A. Calder, J. of Theor. Biology \textbf{100}, 275-282 (1983)

[16] J. Damuth, Nature \textbf{290}, 699-700 (1981)

[17] T. Fenchel, Oecologia \textbf{14}, 317-326 (1974)

[18] B.R. Frieden and R.J. Hughes, Phys. Rev. E \textbf{49}, 2644 (1994).

[19] B. Nikolov and B.R. Frieden, Phys. Rev. E \textbf{49}, 4815 (1994).

[20] A.R. Plastino and A. Plastino, Phys. Rev. E \textbf{54}, 4423 (1996).

[21] B.R. Frieden and W.J. Cocke, Phys. Rev. E \textbf{54}, 257 (1996).

[22] B.R. Frieden, A. Plastino, A.R. Plastino and B.H. Soffer, Phys. Rev. E
\textbf{60}, 48 (1999).

[23] B.R. Frieden, A. Plastino and B.H. Soffer, J. Theor. Biol. \textbf{208},
49 (2001).

[24] R. A. Gatenby and B.R. Frieden, Cancer Res. \textbf{62}, 3675 (2002).

[25] R.J. Hawkins and B.R. Frieden, Physics Lett. A \textbf{322}, 126 (2004)

[26] P.M. Binder, Phys. Rev. E\textbf{61}, R3303 (2000)

[27] R. Badii and A. Politi, $Complexity:\ Hierarchical$ $Structure$ $and$
$Scaling$ $in$ $Physics$ (Cambridge University Press, Cambridge, U.K., 1997)

[28] S. Luo, Found. Phys. \textbf{32}, 757 (2002)

[29] S. Luo, Phys. Rev. Lett. \textbf{91}, 180403 (2003)

[30] B.R. Frieden, $Probability,$ $Statistical$ $Optics$ $and$ $Data$
$Testing,$ $3rd$ $ed.$ (Springer-Verlag, N.Y., 2001)

[31] G.A. Korn and T.M. Korn, $Mathematical$ $Handbook$ $for$
$Scientists\;and$ $~Engineers,~$\ $2nd$ $~ed.$ (McGraw-Hill, N.Y., 1968)

[32] R.V. Churchill, \textit{Fourier Series and Boundary Value Problems}
(McGraw-Hill, N.Y., 1941), particularly pgs. 114-118.

[33] R. A. Gatenby and B.R. Frieden, Math. Biosciences and Eng. \textbf{2},
43-51 (2005)
\end{document}